\documentclass[aps,prl,superscriptaddress,showpacs,preprintnumbers,nofootinbib,reprint,floatfix]{revtex4-1}
\usepackage{mciteplus}
\usepackage{latexsym}
\usepackage{amsthm}
\usepackage{amsmath}
\usepackage{amssymb}
\usepackage{hepunits}
\usepackage{hyperref}
\usepackage{bbm}
\usepackage{bm}
\usepackage{xfrac}
\usepackage{color}
\usepackage{colordvi}
\usepackage{comment}
\usepackage{dcolumn}
\usepackage{times,latexsym,graphicx,wrapfig,url}
\usepackage{epsfig,lineno,bm}
\usepackage{subfigure}
\usepackage{slashed}
\usepackage[normalem]{ulem}
\usepackage{xcolor} 
\usepackage{wasysym}

\usepackage{calrsfs}
\DeclareMathAlphabet{\pazocal}{OMS}{zplm}{m}{n}

\newcommand{\damp}{\eta_{\phi}}

\newcommand{\be}{\begin{eqnarray}}
\newcommand{\ee}{\end{eqnarray}}
\newcommand{\bea}{\begin{eqnarray}}
\newcommand{\eea}{\end{eqnarray}}

\newcommand{\bef}{\begin{figure}[htbp]\begin{center}}
\newcommand{\eef}{\end{center}\end{figure}}
\newcommand\FNAL{Fermi National Accelerator Laboratory, Batavia, IL, 60510, USA}
\newcommand\MIT{Center for Theoretical Physics, Massachusetts Institute of Technology, Cambridge, MA 02139, USA}
\def\lsim{\mathrel{\rlap{\lower4pt\hbox{\hskip1pt$\sim$}}
    \raise1pt\hbox{$<$}}}
\def\gsim{\mathrel{\rlap{\lower4pt\hbox{\hskip1pt$\sim$}}
    \raise1pt\hbox{$>$}}} 

\begin{document}
\preprint{FERMILAB-PUB-17-136-PPD}
\preprint{MIT-CTP/4908}

\title{ Distorted Neutrino Oscillations From Ultralight  Scalar Dark Matter}
\author{Gordan~Krnjaic}          \thanks{krnjaicg@fnal.gov; ORCID: 0000-0001-7420-9577}  \affiliation{\FNAL} 
\author{Pedro~A.~N.~Machado}          \thanks{pmachado@fnal.gov; ORCID: 0000-0002-9118-7354}  \affiliation{\FNAL} 
\author{Lina~Necib}          \thanks{lnecib@mit.edu}  \affiliation{\MIT}

\date{\today}
\begin{abstract}
 Cold, ultralight ($\ll \eV$) bosonic dark matter with a misalignment abundance can induce temporal variation in the masses and couplings of Standard Model particles. We find that fast variations in neutrino oscillation parameters can lead to significantly distorted neutrino oscillations (DiNOs) and yield striking
signatures at long baseline experiments. We study several representative observables to demonstrate this effect and find that current and future experiments including DUNE and JUNO are sensitive to a wide range of viable scalar parameters over many decades in mass reach. 
\end{abstract}

%%%%%%%%%%%%%%%%%%%%%%%%%%%%%%%%%%%%%%%%%%%
\maketitle

\section{Introduction}
%The existence of dark matter (DM) and  neutrino masses are smoking gun evidence of physics beyond the Standard Model (SM); understanding the mechanisms responsible for these phenomena is among the highest priorities in fundamental physics. 

%Ultra light $(\ll \eV)$ bosonic fields are ubiquitous in models of physics beyond the Standard Model (SM). 
%They can arise as axions in solutions to the strong
%CP problem, 

Many popular extensions to the Standard Model (SM) feature ultralight bosonic fields produced non-thermally through the misalignment mechanism.  
In these scenarios, the field is displaced from its minimum in the early universe and begins
to oscillate about this minimum when the Hubble expansion rate becomes comparable to its mass. 
As the universe expands, the field eventually redshifts like nonrelativistic matter
and can account for the dark matter in our universe (for a review see \cite{Essig:2013lka}).

If a misaligned scalar couples to SM fields, it can introduce temporal variation in the measured values of  particle masses and coupling constants.
Recently it has been shown that such time-dependent couplings to neutrinos can affect their masses and mixing angles leading to modulation signals~\cite{Berlin:2016woy}, ameliorate bounds on sterile neutrinos, and modify early universe cosmology~\cite{Berlin:2016bdv,Zhao:2017wmo}.\footnote{Time varying neutrino masses have also been considered 
in the context of dark energy and modified cosmology \cite{Fardon:2003eh,Kaplan:2004dq,Stephenson:1996cz,Hung:2000yg,Kawasaki:1991gn,Stephenson:1996qj,Chacko:2004cz}. }
In this Letter, we show that novel neutrino oscillation signatures arise even when the time variations of neutrino oscillation parameters in a misaligned scalar field background are too fast to lead to any observable modulation.
Even modest couplings to the new field, over a wide range of masses, can significantly modify neutrino oscillation probabilities leading to distorted neutrino oscillations (DiNOs).

To illustrate this effect, consider an ultralight scalar $\phi$ coupled to neutrinos via 
%\be
%\label{eq:leff}
%{\cal L}_{\rm eff} =  -      m_\nu^{ij}  \, \nu_i \nu_j             +  y^{ij}\frac{ \phi }{\Lambda }  \nu_i \nu_j + h.c. ~, 
%\ee
\be
\label{eq:leff}
{\cal L}_{\rm eff} =  -      m_\nu  \left(          1 +        y  \frac{ \phi }{\Lambda }   \right) \nu  \nu  + h.c. ~, 
\ee
where $\Lambda$ is a heavy mass scale and $y$ is a yukawa coupling whose flavor indices have been suppressed.
Such an interaction can arise from a Type-I seesaw model \cite{Minkowski:1977sc,Mohapatra:1979ia} with 
right handed neutrinos $N$ coupled to the scalar field via  
\be
{\cal L } \supset - y_\nu H L N - (\Lambda - y\phi)NN,
\ee
where $L,H$ are the lepton and Higgs doublets, $y_\nu$ and $y$ are Yukawa couplings, and $\phi$ is sequestered from other SM fields.
  Integrating out the $N$ field gives 
the familiar Weinberg operator $(LH)^2/\Lambda$ with an additional $\phi$ interaction, which reduces to Eq.~(\ref{eq:leff}) after electroweak symmetry breaking.
If $\phi$ enjoys an approximate
shift symmetry, it can be naturally light and produced cosmologically through the
misalignment mechanism, thereby constituting some fraction of the dark matter abundance. 

In the present day halo, the local field value can be written 
\be \label{eq:phi-xt}
\phi( x, t) \simeq \frac{\sqrt{2 \rho_\phi^{\tiny\astrosun}}}{~~m_\phi}  \cos[   m_\phi ( t - \vec v \cdot \vec x)    ] ~,
\ee
where $\rho_\phi^{\tiny\astrosun} \le \rho_{\rm DM}^{\tiny \astrosun} = 0.3 \, \GeV\cm^{-3}$  is the scalar energy density and $v\sim 10^{-3}$ is the virial velocity.  
In the presence of this $\phi$ background, the values of neutrino masses are  modulated by the DiNO amplitude  
\begin{equation}
\eta_\phi \equiv \frac{    \sqrt{2\rho_\phi^{\tiny\astrosun}} }{  ~\Lambda m_\phi } ,
\end{equation}
which always refers to the local value of the $\phi$ density.

Note that there are qualitative differences if $\phi$ affects solar parameters ($\theta_{12}, \Delta m^2_{21}$) or atmospheric parameters ($\theta_{13}, \theta_{23}, \Delta m^2_{31}$), as we will see below. Although the impact of such modulation on neutrino mass square differences and mixings depends on the matrix $y$, we analyze the effect of modifying one parameter in isolation assuming $y\sim\pazocal{O}(1)$. 
Thus, each mass squared difference can be written as
\be \label{eq:deltm2}
\Delta m_{ij}^2(x,t) \equiv m_i^2 - m_j^2 \simeq \Delta m_{ij, 0}^2\left(1 +2 \frac{\phi(x,t)}{\Lambda}  \right),
\ee
where $ \Delta m^2_{ij, 0}$ is the undistorted value and $\phi$ evolves according to Eq.~(\ref{eq:phi-xt}),
with corresponding dependence on $\eta_\phi$. Similarly, a potential shift in the mixing angles can be written 
\begin{equation}\label{eq:theta}
  \theta_{ij}(x,t) = \theta_{ij,0}+\frac{\phi(x,t)}{\Lambda},
\end{equation}
where $\theta_{ij,  0}$ is the undistorted mixing angle. 
Note that if the $\phi \nu\nu$ interaction is flavor blind, then the rotations that diagonalize the vacuum mass matrix 
are unaffected and the mixing angles are $\phi$ independent. 
%In this work, we consider the effects in Eqs.~(\ref{eq:deltm2})~and ~(\ref{eq:theta}) in isolation.

In a two-flavor neutrino formalism, the {\it instantaneous} vacuum probability for $\alpha\to \alpha$ survival is
\be \label{eq:prob2flavor}
P(\nu_\alpha \to \nu_\alpha) = 1- \sin^2(2\theta ) \sin^2\left(    \frac{\Delta m^2 L}{4 E}  \right),~~
\ee
where $L$ is the experiment baseline, $E$ is the neutrino energy, and both $\theta$ and $\Delta m^2$ depend on $\phi$ through
Eqs.~(\ref{eq:deltm2}) and (\ref{eq:theta}).
If the scalar oscillation period $\tau_\phi \equiv 2\pi/m_\phi$ is longer than the characteristic neutrino time of flight $T_\nu$, but shorter than the total experimental run time, then neutrinos emitted at different times will sample different values of $\phi$ over the course of a given experiment. 
In this regime, the effective oscillation probability is the ensemble average 
\be \label{eq:average}
\langle  P(\nu_\alpha \to \nu_\beta) \rangle  =
    \int_0^{\tau_\phi} \frac{dt}{\tau_\phi}          P(\nu_\alpha \to \nu_\beta)   ,~~~~~~
\ee 
where for a given experimental baseline $L  = c/T_{\nu}$, there is a characteristic $m_\phi$ 
below which standard  oscillation probabilities can be distorted. 
In Eq.~(\ref {eq:average}) we neglect the spatial variation in $\phi$ since this effect is suppressed by $v \ll 1$. 
If $\tau_\phi \gtrsim 10$ minutes, the misaligned scalar oscillation can induce observable time-variation in neutrino
oscillation measurements (e.g. periodicity in the solar $\nu_e$ flux)~\cite{Berlin:2016woy}. In this work, we study the opposite, high frequency regime and find scalars with $\tau_\phi \ll$ min distort neutrino oscillation probabilities even if this time variation cannot be resolved.

The effect of fast averaging is intrinsically different for neutrino mixing angles and mass-squared differences. For mixing angles, the net effect of averaging over $\phi$ induces a shift in the  observed mixing angle relative to its undistorted value. Note that the observed $\sin^2 2\theta$ after averaging 
can never be zero or maximal since, from Eqs.~(\ref{eq:phi-xt}) and (\ref{eq:theta}), we have
\be 
\label{eq:mixing-smeared}
  \int_0^{\tau_\phi} \frac{dt}{\tau_\phi}\sin^2 2\theta(t) &=&  \frac{1}{2} \bigl[1 - J_0(4\eta_\phi)  \cos 4\theta_0   \bigr]  \\
  &\simeq& \sin^2 2\theta_0\left(1-4\damp^2\right) +2\damp^2  + {\pazocal O}(\eta_\phi^3),~~~~~~~~~
\ee
where $J_0$ is a Bessel function of the first kind and, to quadratic order in $\eta_\phi$, the correction to the $\sin^22\theta(t)$ distribution is negative (positive) for maximal (minimal) mixing. 
Thus the observations of non-zero $\theta_{13}$~\cite{An:2016ses} and nearly 
maximal $\theta_{23}$~\cite{Abe:2017uxa, Adamson:2017gxd} already constrain the available parameter space.

\begin{figure}[t!] 
\hspace{-0.cm}
\includegraphics[width=9cm]{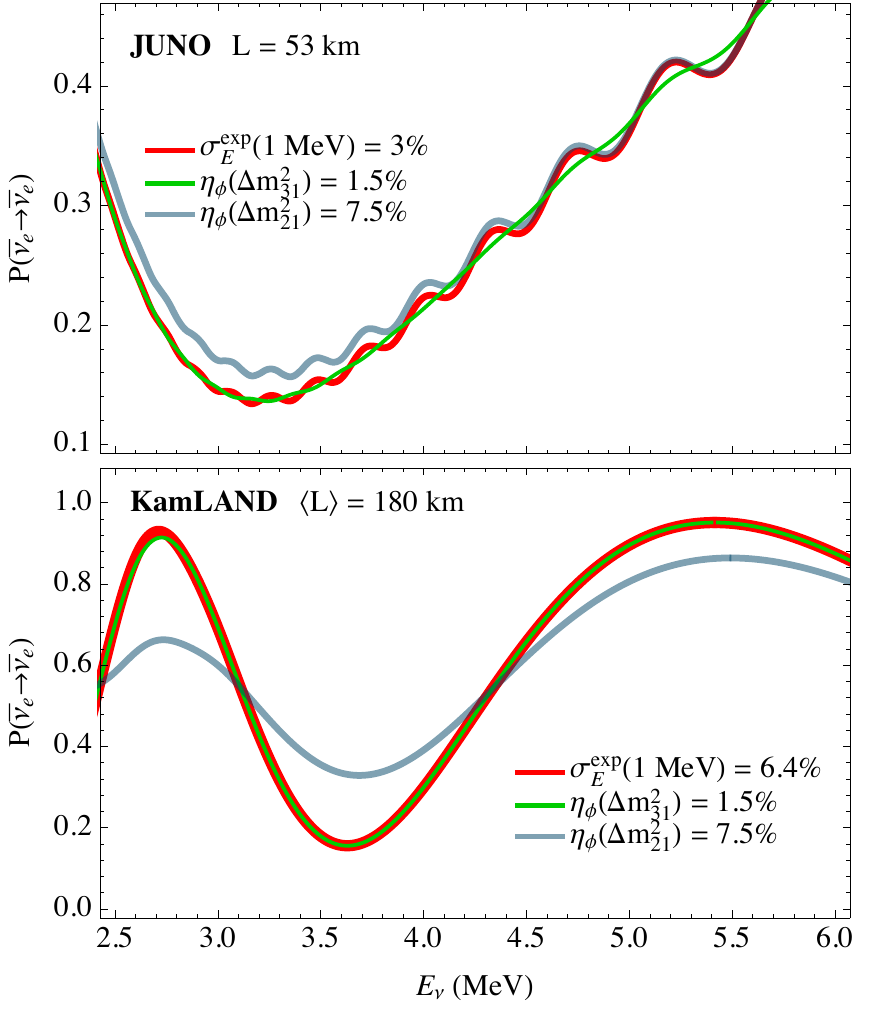}  ~~~~~\vspace{-0.7cm}
   \caption{Example neutrino oscillation probabilities for a variety of scenarios at JUNO  (top) and KamLAND (bottom). For both plots, the thick red curve is the standard oscillation prediction for each setup including the effect of energy resolution smearing (following the prescription in Appendix A). The green and turquoise curves also include the additional effect of $\phi$-induced smearing 
   separately distorting  $\Delta m^2_{31}$  and $\Delta m^2_{21}$, respectively.  For KanLAND we have assumed a mean baseline between the nuclear reactors and the detector of $\langle L\rangle=180$~km.\label{fig:JUNO}           }
\vspace{0cm}
\end{figure}

If the scalar primarily affects mass-squared differences (e.g. through flavor blind yukawa couplings), the time averaging has a more 
complicated functional dependence  
\be\label{eq:average-mass-sq}
 \int_0^{\tau_\phi} \frac{dt}{\tau_\phi} \sin^2\left[ \frac{\Delta m^2 L }{4 E} \left(1 +  2\eta_\phi \cos m_\phi t       \right)   \right],~~~
\ee
which leads to additional $L/E$ smearing and distorts the functional form of oscillation probabilities, particularly near maxima and minima. 
Thus, the DiNO effect from Eq.~(\ref{eq:average}) adds an irreducible 
smearing to the oscillation probability signal, similar to an experimental energy resolution,
but at the probability level. This effect is shown in Figs.~\ref{fig:JUNO} and \ref{fig:DUNE}, which present both instantaneous and $\phi$-averaged $\overline{\nu}_e \rightarrow \overline{\nu}_e$ survival probabilities as a function of neutrino energy for
JUNO~\cite{An:2015jdp} and KamLAND \cite{Eguchi:2002dm,Araki:2004mb} 
as well as $\nu_\mu \rightarrow \nu_\mu$ and  $\nu_\mu \rightarrow \nu_e$ oscillation probabilities at the future experiment DUNE~\cite{Acciarri:2015uup} (see Appendix A for a discussion of the signal calculation).

\section{Phenomenology} 
\label{sec:sensitivity}

\begin{figure*}[t!] 
\includegraphics[width=17.5cm]{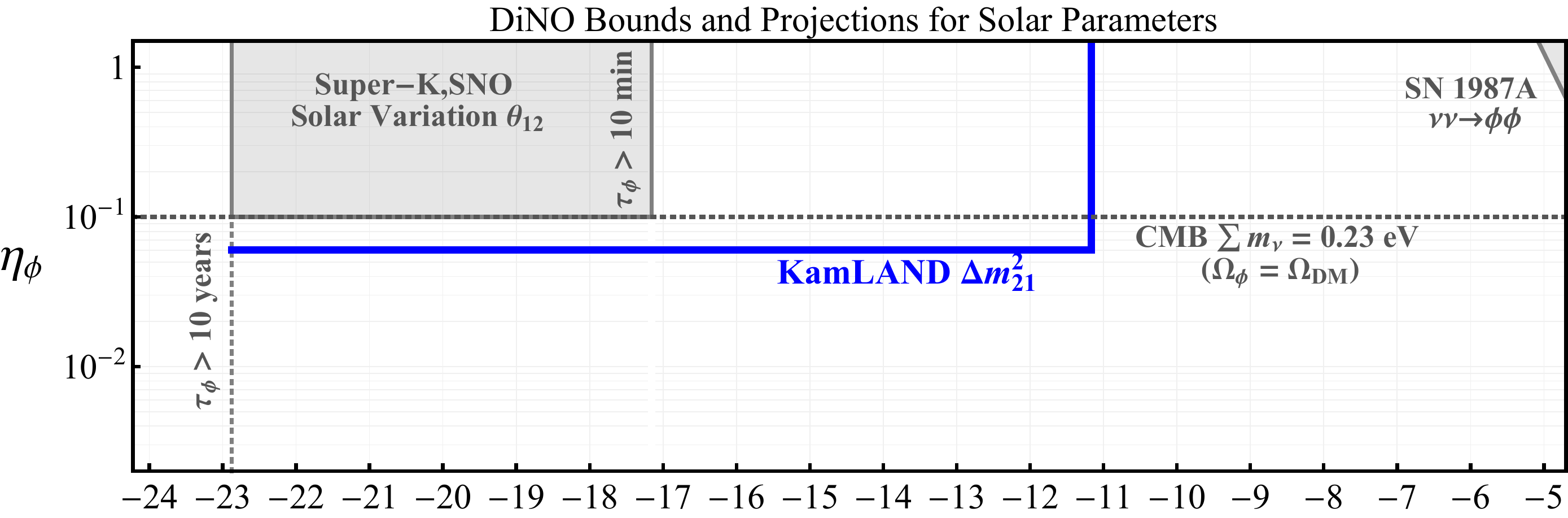}  ~~~~~
\vspace{-0.5cm}
\includegraphics[width=17.5cm]{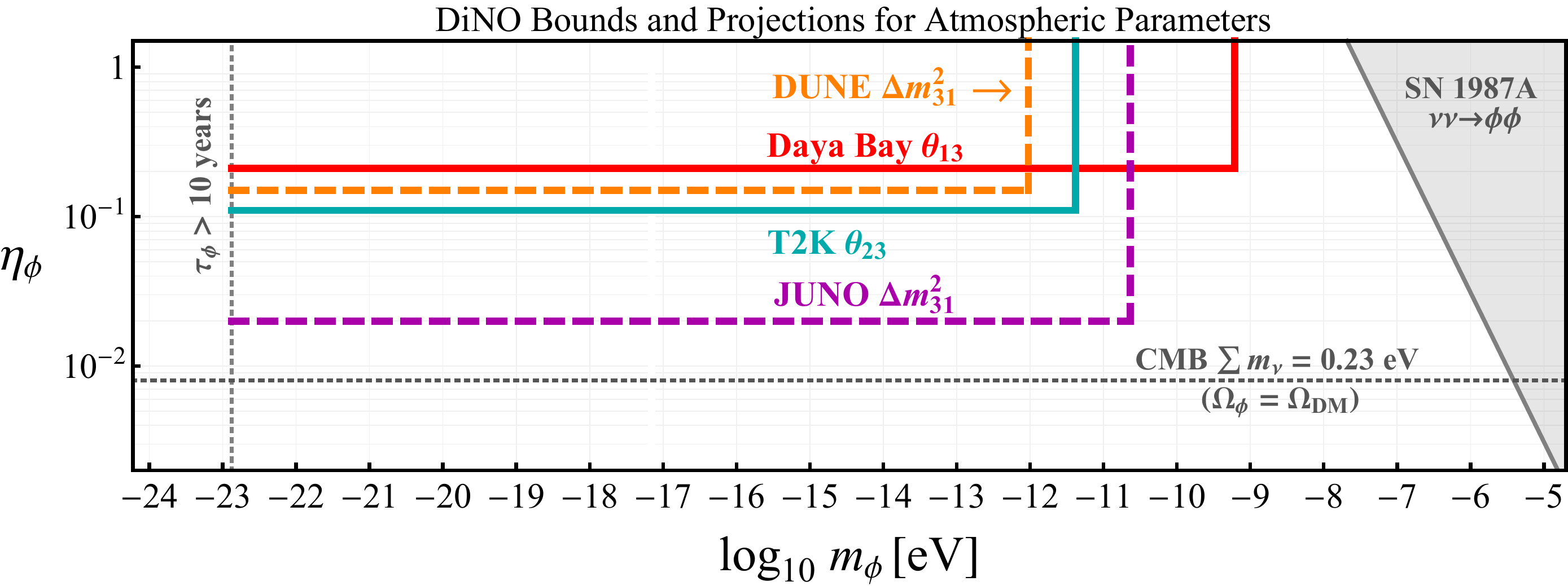}  ~~~~~
\caption{
Viable parameter space for $\phi$-induced variations in the neutrino mass matrix in terms of $\damp$ assuming that $\phi$ affects 
only solar oscillation parameters (top panel) or only atmospheric oscillation parameters (bottom panel). The upper left region labeled ``$\tau_\phi >$ 10 min'' is the bound  
on anomalous periodicity in solar neutrino oscillations at  Super-K/SNO \cite{Berlin:2016woy} and the leftmost region labeled ``$\tau_\phi >10$ years'' corresponds
to scalar periods that are too long to have an observable effect in terrestrial experiments. In this regime, $\phi$ is effectively a constant background contribution to neutrino masses and mixings.
 The diagonal shaded region in both plots is the bound from energy loss in SN1987A (see text and Appendix B). The dotted gray lines are the CMB bound 
 on $\sum_i m_{\nu_i} < 0.23~\eV$ if $\phi$ constitutes all of the dark matter at the time of recombination; the line moves between plots
 since the effect of coupling to light eigenstates (top panel) has less of an overall effect (see text). The solid cyan (red) line labeled
 ``T2K $\theta_{23}$'' (``Daya Bay $\theta_{13}$'') indicates the current exclusion due to near maximal 
 (minimal) measurement of $\theta_{23}$ ($\theta_{13}$).  Similarly, the solid blue line labeled ``KamLAND $\Delta m^2_{21}$'' is the existing bound from $\Delta m^2_{21}$ $\phi$-smearing.
  The dashed orange and purple lines indicate the projected sensitivities for 
 DUNE and JUNO, respectively. Note that each constraint depends on time variations only on individual parameters as labeled (see text for details). 
 }
   \label{fig:MoneyPlot}
\vspace{0cm}
\end{figure*}

Although a detailed experimental analysis is outside the scope of the paper,\footnote{Such an analysis would require a careful treatment of neutrino energy reconstructions, a daunting task to anyone outside the experimental collaborations.} we present estimates of the experimental sensitivities of current and future neutrino experiments in terms of the ratio $\damp$  taking into account possible new interpretations of oscillation parameters. In Fig.~\ref{fig:MoneyPlot} we summarize our main results as bounds and projections on the $m_\phi-\eta_\phi$ plane assuming separately that $\phi$ only affects solar (top panel) and atmospheric oscillations (bottom panel).

\begin{figure}[t] 
\hspace{-0.cm}
\includegraphics[width=8.8cm]{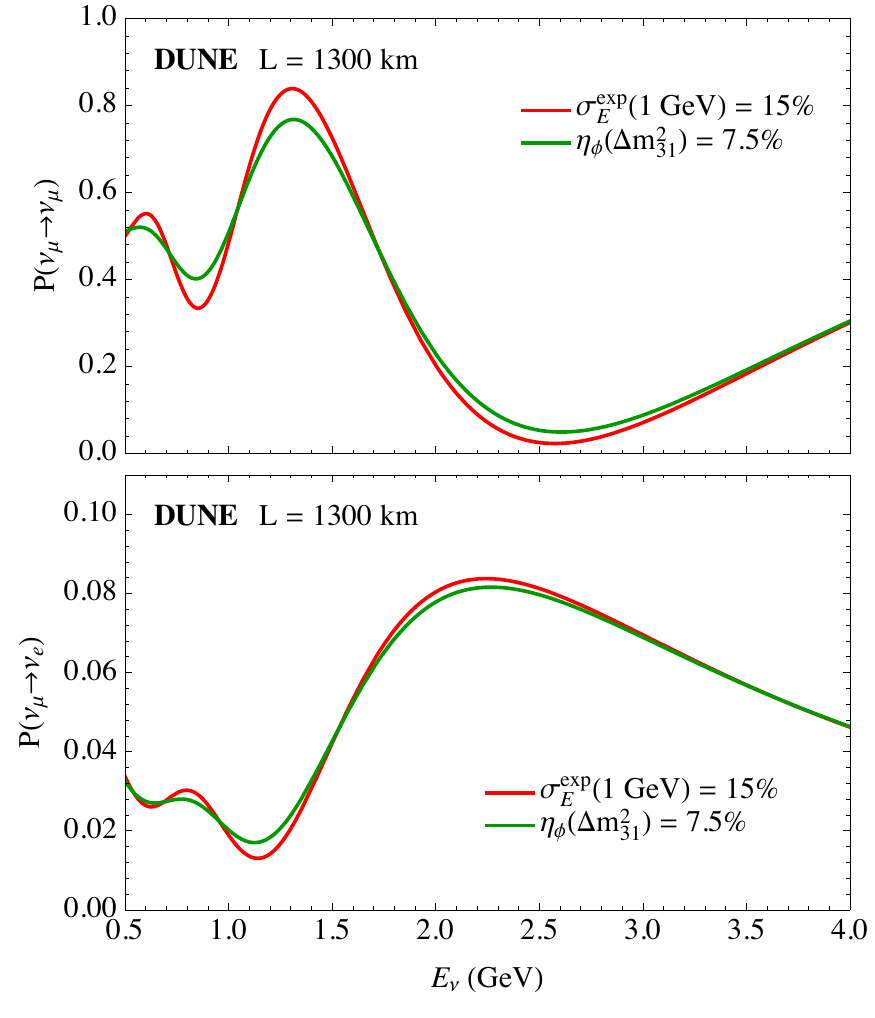}
   \caption{Effect of $\phi$-induced distortion on DUNE disappearance (top) and appearance (bottom) oscillation probabilities through $\Delta m^2_{31}$ smearing.}
   \label{fig:DUNE}
\vspace{0cm}
\end{figure}

\begin{itemize}
\item{\bf Bounds from $\Delta m^2$:}
 The presence of a time-varying $\phi$ background influences the interpretation of existing
 neutrino oscillation parameters in Eq.~(\ref{eq:mixing-smeared}).
 The excellent energy resolution of the KamLAND experiment of $6.4\%$ at 1~MeV~\cite{Gando:2013nba} constrains $\damp\lesssim 3.2\%$ via $\Delta m^2_{21}$ smearing (e.g. induced by nonzero $y_{11}$ in Eq.~\ref{eq:leff}), as shown in Fig.~\ref{fig:JUNO}, where we illustrate the effect of distorting $\Delta m_{31}^2$  (e.g. induced by nonzero $y_{33}$ in Eq.~\ref{eq:leff}) and $\Delta m_{21}^2$ on the disappearance probability of $\overline{\nu}_e$. Note that the dip in the oscillation minimum for KamLAND is set by  $\theta_{12}$, in analogy with Eq.~(\ref{eq:prob2flavor}). However, the heights of the maxima are affected by $\theta_{13}$ induced atmospheric oscillations (which are smeared out due to the experimental resolution), and thus are essentially fixed due to the current precision of $\theta_{13}$ measurements. DiNOs therefore distinctively affect the height of the maxima in KamLAND. 
 
Dedicated analyses are required to estimate the sensitivity to $\phi$-smearing in $\Delta m^2_{31}$ distortion at current experiments: at MINOS~\cite{Adamson:2014vgd} the  neutrino flux  peaks away from  the oscillation minimum; Super-K atmospheric data~\cite{Fukuda:1998mi} has a non-trivial neutrino flux; and both NO$\nu$A \cite{Ayres:2004js} and T2K \cite{Abe:2011sj} are narrow band beams which do not resolve the functional form of the oscillation probability.

 \item{\bf Bounds from mixing angles:}
 The T2K observation of $\sin^2\theta_{23}=0.53^{+0.08}_{-0.11}$~\cite{Abe:2017uxa} constrains
 $\damp<0.11$, see Eq.~(\ref{eq:mixing-smeared}), while the current value of $\sin^2 2\theta_{13}=0.0841\pm0.0033$ measured by Daya Bay~\cite{An:2016ses} provides a weaker constraint $\damp<0.21$.

\item{\bf Future neutrino oscillation experiments:} 
The future JUNO experiment aims to measure the neutrino hierarchy by observing the small-amplitude, high frequency oscillations caused by $\Delta m^2_{31}$ in solar baseline oscillations. Its proposed energy resolution of $3\%$  translates into a sensitivity of $\damp\sim1.5\%$, as shown in Fig.~\ref{fig:JUNO}, where we illustrate the disappearance probability $P(\overline{\nu}_e \rightarrow \overline{\nu}_e)$ in JUNO for $\phi$-smearing effects on $\Delta m_{21}^2$ and $\Delta m_{31}^2$. The effect on $\Delta m^2_{21}$, although non-negligible, can be mimicked by adjusting the undistorted value of $\theta_{12}$, as we verified numerically. The DUNE experiment will also provide a constraint on $\damp$ of the order of half of its energy resolution (around 15\% at 1~GeV) due to $\Delta m^2_{31}$ smearing as shown in Fig.~\ref{fig:DUNE}, where we illustrate the $\phi$-smearing effect on $\Delta m_{31}^2$ on the disappearing probability $P(\nu_\mu \rightarrow \nu_\mu)$ as well as the appearance probability $P(\nu_\mu \rightarrow \nu_e)$. The effect of $\phi$-smearing on $\Delta m^2_{21}$ for DUNE is negligible. Future determinations of $\theta_{23}$ by NO$\nu$A \cite{Adamson:2017gxd}, T2K \cite{Batkiewicz:2017xoh}, DUNE \cite{Acciarri:2015uup}, and T2HK \cite{Ishida:2013kba} can bound $\damp$ \emph{if the observed of $\theta_{23}$  is  maximal}. 
\end{itemize}

%
%\section{Existing Constraints}
%\label{sec:constraints}

%We present the existing constraints on $\phi/\Lambda$ in fig.~\ref{fig:MoneyPlot}. Below follows the explanation of each bound.

\begin{figure}[t!] 
~~~\includegraphics[width=7.6cm]{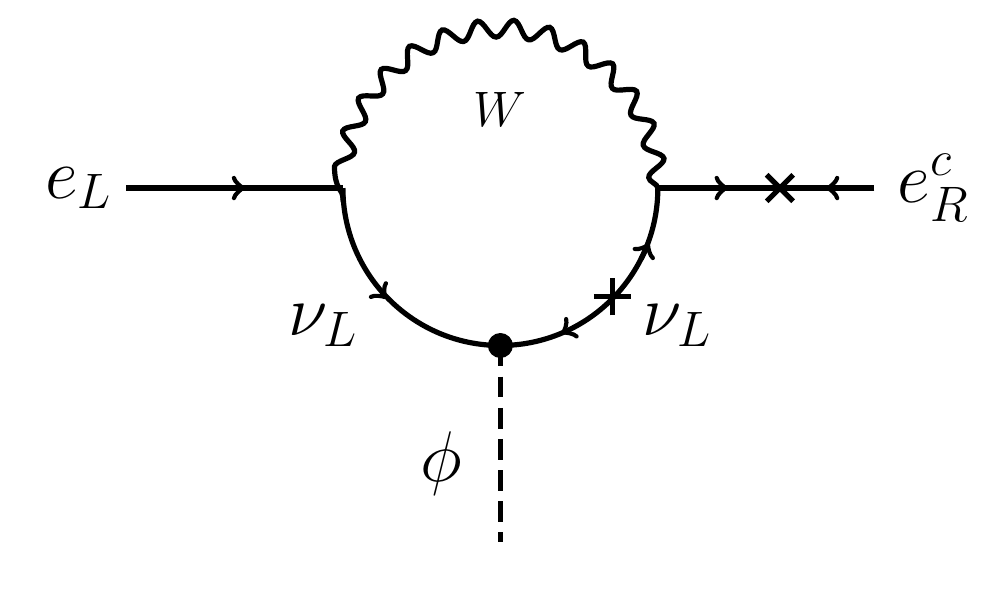}  ~~~~~
\caption{Leading diagram that contributes to electron mass variation due to oscillating scalar couplings at loop level. Due to
multiple neutrino mass insertions, the amplitude for this fractional variation is on the order of $\sim 10^{-29}$ (see text for a discussion).}
   \label{fig:feynman-electron-loop}
\vspace{0cm}
\end{figure}

\begin{itemize} 
\item{\bf CMB:} 
Since the average cosmological $\phi$ density redshifts as nonrelativistic matter, $\rho_{\phi}(z) \propto (1+z)^{3}$, where $z$ is the redshift, the amplitude of neutrino mass modulation is much larger at earlier times. 
As observed in Ref. \cite{Berlin:2016woy}, a sharp increase in the overall scale of neutrino masses in the early universe can exceed the Planck limit on their sum $\sum_i m_{\nu_i} < 0.23~ \eV$ \cite{Ade:2015xua}.  For concreteness, in this discussion we assume a normal neutrino mass ordering. 

{\bf $\phi$ coupled to heavy eigenstate:}  If the $\phi$ affects the heaviest neutrino, the absolute neutrino mass scale is modified and we demand the average $\phi$-induced correction be no greater than an $\pazocal{O} (1)$ 
effect at recombination
\be \label{eq:sum-cmb}
 ~~~~~~\eta_\phi (z_{\rm rec}) =  \frac{\sqrt{2 \rho_{\phi}(z=0) }}{ \Lambda m_\phi}   \left(   1+z_{\rm rec}    \right)^{3/2} \lesssim 1,~~~~
\ee
where $z_{\rm rec} \simeq 1100$ is the redshift at recombination, $ \rho_{\phi}(z=0) $ is the present day average {\it cosmological}  $\phi$ density, and 
we have  taken $m_\nu \sim 0.1~ \eV$. 
 Translating this into  a bound on the local modulation amplitude in the solar neighborhood, we find 
\be \label{eq:sum-cmb-present}
\eta_\phi(z = 0)   \equiv  \frac{\sqrt{2 \rho_{\phi }^{\tiny\astrosun}}}{ ~~\Lambda m_\phi}  \lesssim      9 \times 10^{-3}~~,
\ee
where  and $\rho_\phi^{\tiny\astrosun} $ is the \textit{the local $\phi$ density} and we have assumed the usual halo overdensity relation $\rho_\phi^{\tiny\astrosun}  \sim 10^{5} \rho_{\phi}(z=0)$. 

{\bf $\phi$ coupled to lighter eigenstates:} In this regime, a similar argument applies, but $\phi$ now couples only to light neutrinos with $m_1 \sim m_2 \sim  \sqrt{\Delta m^2_{21}} \approx 0.008$ eV. This assumption translates into the requirement  
\be \label{eq:sum-cmb-present-light}
\eta_\phi(z = 0)   \equiv  \frac{\sqrt{2 \rho_{\phi }^{\tiny\astrosun}}}{ ~~\Lambda m_\phi}  \lesssim    0.1 ~~.
\ee

However, the bounds in Eqs.~(\ref{eq:sum-cmb-present}) and (\ref{eq:sum-cmb-present-light}) apply only if $\phi$ accounts for all of the dark matter at recombination; if it only constitutes  
a subdominant fraction of the DM density,  it need not be dynamical in the early universe, so the constraint no longer applies.
In this work, wherever the modulation effect exceeds these bounds, we will assume that $\phi$ oscillation begins after recombination.

%%%%%%%%%%%%%%%%%%%%%%%%%%%%%%%%%%%%%%%%%%%%
%					Solar Oscillation Periodicity
%%%%%%%%%%%%%%%%%%%%%%%%%%%%%%%%%%%%%%%%%%%%
\item{\bf Solar neutrino periodicity:} The observed temporal stability of solar neutrino fluxes by Super-K imposes
a tight bound on neutrino mass variation over 10 min--10 year timescales \cite{Fukuda:2001nj,Berlin:2016woy}.  The period of 
$\phi$ induced mass variation in our setup is 
\be \label{eq:period}
\hspace{0.3cm} \tau_\phi = \frac{2\pi}{~m_\phi}  \simeq  10\, {\rm min} \left( \frac{7 \times 10^{-18} \, \eV }{m_\phi}  \right),
\ee
so as long as  $m_\phi \gtrsim 10^{-17} \eV$, the period is too short to have been observed at Super-K. If, instead,
the period is within the Super-K time resolution, the $\eta_\phi$ amplitude is bounded to be below $\pazocal{O}(10\%)$~\cite{Berlin:2016woy}. 

%%%%%%%%%%%%%%%%%%%%%%%%%%%%%%%%%%%%%%%%%%%%
%					Supernova 1987a
%%%%%%%%%%%%%%%%%%%%%%%%%%%%%%%%%%%%%%%%%%%%
\item{\bf SN 1987A:}
  Light weakly coupled particles can introduce anomalous energy loss in supernova (SN), thereby 
conflicting with observations from SN1987A \cite{Hirata:1987hu,Bionta:1987qt}. Since $\phi$ 
only couples to neutrinos, the total energy released due to $\nu \nu \to \phi \phi$ annihilation for a SN temperature of $T$ = 30 MeV is    
\be
~~~~~\Delta E_{\rm \phi} \sim 4 \times 10^{50} \, {\rm erg}  \, \left( \frac{ 50\, \keV}{\Lambda}\right)^4     \left( \frac{ \Delta t}{10 \, \sec }\right),~~
\ee
where $\Delta t $ is the SN blast duration. 
The observed energy released in SN1987A is approximately $10^{51}$ erg, so to avoid an order one correction, we demand $\Lambda \gtrsim 50 ~\keV$ (see Appendix B for details). 
 
 However, the  neutrino mass modulation amplitude scales as $\damp$, so fixing the magnitude of this effect
 and saturating the SN bound on $\Lambda \sim 50 ~\keV$ implies a maximum scalar mass
 \be
 \hspace{0.5cm} ~~~~~m_{\phi} \lesssim  3 \times 10^{-7} \, \eV   
    \left(    \frac{ 0.1}{\eta_\phi} \right)                   \left(    \frac{ 0.1 \,  {\rm eV}  }{m_{\nu,i}} \right)^2  \!
    \left(  \frac{ \Omega_{\phi} }{\Omega_{\rm DM}} \right)^{  \!\! 1/2}\!\!\!\!,~~~~~~~~~
 \ee
where the index in $m_{\nu, i}$ refers to heaviest neutrino mass eigenstate to which $\phi$ couples. This bound defines the diagonal shaded region in Fig. \ref{fig:MoneyPlot} where we take $\Omega_\phi  = \Omega_{\rm DM}$. 

%%%%%%%%%%%%%%%%%%%%%%%%%%%%%%%%%%%%%%%%%%%%
%					Charged Lepton Variation
%%%%%%%%%%%%%%%%%%%%%%%%%%%%%%%%%%%%%%%%%%%%
\item{\bf Charged Lepton Mass Variation:} 
A time varying neutrino mass can induce charged lepton mass variation through the loop level diagram depicted in Fig. \ref{fig:feynman-electron-loop}. However,
the amplitude of this variation is estimated to be  
\be
~~~~~~~~~~\frac{\delta m_e}{m_e} \sim \frac{G_F \, m_\nu^2 }{16 \pi^2} \, \eta_\phi \! \sim  10^{-28} \!  \left(\frac{\eta_\phi }{ \, 0.1 \,}\right)    
 \! ,~~~~~
\ee
where we have taken $m_\nu = 0.1\, \eV$. Compared to the  existing limit on electron mass time-variation $\delta m_e / m_e \lesssim 10^{-16}$ from the stability of atomic clocks \cite{Luo:2011cf}, this effect is 
negligible. 
However, if Higgs portal operators of the form $\phi H^\dagger H$ and $\phi^2 H^\dagger H$ are allowed, there will be additional time variation in fermion
masses due to Higgs mixing after electroweak symmetry breaking. The effect of these operators can be naturally suppressed, for instance, by extra dimensional locality (e.g. if $\phi$ and $H$ live on different branes in a higher dimensional model), however the model building details are beyond the scope of this work.
\end{itemize}

%%%%%%%%%%%%%%%%%%%%%%%%%%%%%%%%%%%%%%%%%%%%
%					Existing Bounds From Oscillations 
%%%%%%%%%%%%%%%%%%%%%%%%%%%%%%%%%%%%%%%%%%%%

\section{Conclusion}
In this Letter we have found that ultra-light scalar dark matter can significantly distort neutrino oscillation phenomenology by introducing rapid time variation in neutrino masses and mixing angles. As neutrinos travel between source and detector, they traverse a spacetime dependent scalar background, which
must be averaged over when calculating flavor transition probabilities. 
If this effect primarily modifies mixing angles, it can shift the extracted parameters away from extremal values (maximal or minimal mixing), thereby 
requiring a reinterpretation of $\theta_{13}$ and $\theta_{23}$ measurements. 
Alternatively, if the scalar modulation primarily affects mass-squared differences, it can distort the functional form of 
oscillation probabilities (e.g. the locations of maxima/minima) akin to an additional energy resolution effect. Thus, experiments with good energy 
reconstruction (e.g. KamLAND, JUNO and DUNE) are particularly sensitive to this scenario.
 
We find that existing neutrino oscillations experiments, including KamLAND, T2K, and  Daya Bay  already exclude a large portion of the scalar parameter space, covering over 
a dozen orders of magnitude in mass reach; however, the applicability of these limits depends on whether  the misaligned scalar modifies the parameters extracted at each experiment. Future experiments including JUNO and DUNE are poised  to significantly extend this coverage.

%%%%%%%%%%%%%%%%%%%%%%%%%%%%%%%
\begin{acknowledgments}
{\it Acknowledgments} We thank Asher Berlin, Cliff Burgess, Pilar Coloma, Yuval Grossman, Roni Harnik, Christoper T. Hill, Kiel Howe, Brian Shuve, Jesse Thaler, and Felix Yu for helpful discussions.  Fermilab is operated by Fermi Research Alliance, LLC, under Contract No. DE- AC02-07CH11359 with the US Department of Energy.   L.N. is supported by the DOE under contract DESC00012567. P.M. acknowledges partial support from the EU grants H2020-MSCA-ITN-2015/674896-Elusives and H2020-MSCA-2015-690575-InvisiblesPlus.
\end{acknowledgments} \medskip

\bibliography{NeutrinoScalars}
%\newpage

%\onecolumngrid

%\newpage
\appendix
\setcounter{equation}{0}
\setcounter{figure}{0}
\setcounter{table}{0}
\setcounter{section}{0}
\makeatletter
\renewcommand{\theequation}{A\arabic{equation}}
\renewcommand{\thefigure}{A\arabic{figure}}
\renewcommand{\thetable}{A\arabic{table}}

\section{Supplementary Material}

\appendix

\section{Appendix A: Details of Oscillation Averaging}
\label{sec:signal-appendix}

In this appendix we present the full procedure for calculating three flavor neutrino oscillation observables with energy resolution smearing. For neutrino flavor transitions undistorted by $\phi$, the two flavor probability in Eq.~(\ref{eq:prob2flavor})  generalizes to
 \be
\hspace{-0.1cm} P(\nu_\alpha \!\! \to \! \nu_\beta) &=& \delta_{\alpha\beta}  - 4 \! \sum_{i>j} {\Re}(U^{*}_{\alpha i}   U_{\beta i}   U_{\alpha j}U^{*}_{\beta j} )  \sin^2 \! \left(  \!  \frac{\Delta m^2_{ij} L}{4E}   \!  \right)  \nonumber \\ 
&&  +2 \sum_{i>j} {\Im}(U^{*}_{\alpha i}   U_{\beta i}   U_{\alpha j}U^{*}_{\beta j} ) \sin\left(\frac{\Delta m^2_{ij} L}{4E}\right),~~~~~~~~~
\ee
where $U$ is the PMNS matrix (in the presence of matter effects~\cite{Wolfenstein:1977ue,Mikheev:1986gs}), $\alpha, \beta = e,\mu,\tau$ and $i,j = 1-3$ and the neutrino oscillation parameters are given in ref.~\cite{Agashe:2014kda}. 
All of our numerical results adopt this ansatz for the standard oscillation probability for the integrand in Eq.~(\ref{eq:average}). Although we include the MSW effect \cite{Wolfenstein:1977ue,Mikheev:1986gs} on our calculation of the oscillation probabilities,  it does not differentially modify oscillations in the presence of the $\phi$ background.

  To model  the experimental energy resolution we evaluate the effective probability 
  \be
  P(E_{\rm r})=\int_0^\infty  dE_{\rm t}P(E_{\rm t}) f(E_{\rm t}, E_{\rm r}, \sigma_E^{\rm exp}),
  \ee
  where $f$ is a Gaussian distribution modeling the energy reconstruction, namely
  \begin{equation}
    f \propto \exp\left[-\frac{1}{2}\left(\frac{E_{\rm t}- E_{\rm r}}{E_{\rm t}\sigma }\right)^2\right],
  \end{equation}
with  $\sigma = \sigma_E^{\rm exp} /\sqrt{E_{\rm t}/E_0}$ is the energy resolution at a reference energy $E_0$, and $E_{\rm t,r}$ are the true and reconstructed energies. Note that here $f$ has been normalized such that $\int_0^\infty dE_{\rm t} f(E_{\rm t}, E_{\rm r}, \sigma_E^{\rm exp}) =1.$  

Adding the effect of $\phi$-averaging to the oscillation probability yields
\begin{equation}
  \langle  P(E_{\rm r}) \rangle  =
    \int_0^\infty  dE_{\rm t}\int_0^T \frac{dt}{T}    P(E_{\rm t}) f(E_{\rm t}, E_{\rm r}, \sigma_E^{\rm exp}),
\end{equation}
which defines the procedure for obtaining the oscillation curves in Figs.~\ref{fig:JUNO} and \ref{fig:DUNE}.

\appendix
\setcounter{equation}{0}
\setcounter{figure}{0}
\setcounter{table}{0}
\setcounter{section}{0}
\makeatletter
\renewcommand{\theequation}{B\arabic{equation}}
\renewcommand{\thefigure}{B\arabic{figure}}
\renewcommand{\thetable}{B\arabic{table}}

\appendix
\setcounter{equation}{0}
\setcounter{figure}{0}
\setcounter{table}{0}
\setcounter{section}{0}
\makeatletter
\renewcommand{\theequation}{B\arabic{equation}}
\renewcommand{\thefigure}{B\arabic{figure}}
\renewcommand{\thetable}{B\arabic{table}}

\section{Appendix B: Supernova Bounds}
In this appendix we derive an order-of-magnitude bound on $\eta_\phi$ from SN1987a by making some conservative assumptions on the emission rate of $\phi$ from the 
thermalized core of neutrinos inside the supernova with $T \sim 30 \, \MeV$. The cross section for $\phi$ emission via  $\nu \nu \to  \phi\phi$ 
annihilation is estimated to be  
\be
\sigma(\nu \nu \to \phi \phi) \sim \frac{1}{T^2} \left( \frac{m_\nu}{\Lambda}\right)^4  ,
\ee
where we adopt $m_\nu = 0.1 \, \eV$. The $\phi$ emission rate 
per neutrino in a  thermal bath with number density $n_\nu =  9\zeta(3) T^3/4 \pi^2 =  7\times10^{-6} \, \GeV^3$ is
\be
\dot E_\nu \sim T n_\nu \sigma v \sim  4 \times 10^{-27} \, \GeV^2 \,    \left( \frac{ 50\,  \keV}{\Lambda}\right)^4 .~~
\ee
For a SN radius of $R \sim 10 \, {\rm km}$, we have approximately $N_\nu \sim 4 \pi R^3  n_\nu/3  \sim 4 \times 10^{54} $ neutrinos
in the core, which implies an anomalous cooling rate of
\be
\dot E_{\phi} \sim \dot E_\nu N_{\nu}    \sim ~   1.5 \times 10^{28} \, \GeV^2 \,    \left( \frac{50\, \keV}{\Lambda}\right)^4,~~~~~~~~
\ee
so the total energy loss per $\Delta t = 10 \sec$  burst is 
\be
\Delta E_{\rm \phi}  \sim 4 \times 10^{50} \, {\rm erg}  \, \left( \frac{ 50\, \keV}{\Lambda}\right)^4     \left( \frac{ \Delta t}{10 \, \sec }\right).~~~~~~~~
\ee
To avoid adding an order one correction to the energy loss from SN1987a, we demand  $ \Delta E_{\phi} \lesssim 10^{51}  \, {\rm erg}$, which imposes the modest limit $ \Lambda \gsim   50~\keV$.  Since most of the couplings
we consider in this work are vastly greater than this bound, this bound imposes essentially no  limit on the relevant parameter space, as shown in the bottom panel of Fig.~\ref{fig:MoneyPlot}.  Note that in the top panel, the SN bound assumes that $\phi$ only affects solar oscillation parameters and, therefore, only couples 
to the lighter neutrino mass eigenstates. Since the cross section scales as $\sigma(\nu \nu \to \phi \phi) \propto m^2_\nu$ and we now take the $m_1 \sim m_2 \sim 0.005$ eV as a reference mass scale, the bound is correspondingly weaker.

\end{document}